\documentclass[english,aps,a4]{revtex4}
\usepackage[T1]{fontenc}
\usepackage[latin9]{inputenc}
\usepackage{color}
\usepackage{amsmath}
\usepackage{amssymb}
\usepackage{graphicx}
\usepackage{esint}

\makeatletter
\@ifundefined{textcolor}{}
{%
 \definecolor{BLACK}{gray}{0}
 \definecolor{WHITE}{gray}{1}
 \definecolor{RED}{rgb}{1,0,0}
 \definecolor{GREEN}{rgb}{0,1,0}
 \definecolor{BLUE}{rgb}{0,0,1}
 \definecolor{CYAN}{cmyk}{1,0,0,0}
 \definecolor{MAGENTA}{cmyk}{0,1,0,0}
 \definecolor{YELLOW}{cmyk}{0,0,1,0}
}

\usepackage{subfigure}

\allowdisplaybreaks[1]

\usepackage{babel}

\makeatother

\usepackage{babel}
\begin{document}

\title{Instability in a minimal bimetric gravity model}

\author{Frank Koennig$^{1}$, Luca Amendola$^{1}$}

\affiliation{$^{1}$Institut Für Theoretische Physik, Ruprecht-Karls-Universität
Heidelberg, Philosophenweg 16, 69120 Heidelberg, Germany}
\begin{abstract}
We discuss in detail a particularly simple example of a bimetric massive
gravity model which seems to offer an alternative to the standard
cosmological model at background level. For small redshifts, its equation
of state is $w(z)\approx-1.22_{-0.02}^{+0.02}-0.64_{-0.04}^{+0.05}z/(1+z)$.
Just like $\Lambda$CDM, it depends on a single parameter, has an
analytical background expansion law and fits the expansion cosmological
data well. However, confirming previous results, we find that the
model is unstable at early times at small scales and speculate over
possible ways to cure the instability. In the regime in which the
model is stable, we find that it fits the linear perturbation observations
well and has a growth index of approximately $\gamma=0.47$. 
\end{abstract}
\maketitle

\section{\textcolor{black}{Introduction}}

\textcolor{black}{The history of massive gravity dates back to 1939,
when the linear model of Fierz and Pauli was published (see e.g. Refs.
\cite{2012RvMP...84..671H} and \cite{2014arXiv1401.4173D} for a
review). Massive gravity requires the introduction of a second tensor
field in addition to the metric (or some form of nonlocality in the
action; see Ref. \cite{Maggiore:2014sia}). The interaction of the
two tensor fields creates a mixture of massless and massive gravitons
that apparently avoids the appearance of ghosts \cite{2010PhLB..693..334D,2010PhRvD..82d4020D,2011PhRvD..84d3503D,2011PhRvL.106w1101D}.}

\textcolor{black}{In the model introduced in Refs. \cite{2012JHEP...02..126H,2012JHEP...02..026H},
the second tensor field becomes dy}namical, just like the standard
metric, although only the latter is coupled to matter (for a generalization,
see \textcolor{black}{Ref.} \cite{2013JCAP...10..046A}). This approach,
denoted bimetric gravity, keeps the theory ghosts free and has the
advantage of allowing cosmologically viable solutions. The cosmology
of bimetric gravity has been studied in several papers, e.g. in Refs.
\cite{1475-7516-2012-03-042,2013JHEP...03..099A,Comelli:2012db,2013arXiv1304.3920D,2012JHEP...03..067C,2012JHEP...01..035V,2012JCAP...12..021B}.

In this paper we select among the class of bimetric models a particularly
simple case, which we dub the minimal bimetric model (MBM). Just like
$\Lambda$CDM, this model depends on a single parameter and has an
analytical background behavior that is at all times distinguishable
from $\Lambda$CDM. In a previous paper we have shown that the MBM
is the only one-parameter version of bimetric gravity (beside the
trivial case in which only a cosmological constant is left) that is
cosmologically well behaved at the background level and fits the supernovae
Hubble diagram well \cite{1475-7516-2014-03-029} (see also Refs.
\cite{1475-7516-2012-03-042,2013JHEP...03..099A}).

Unfortunately, considering the full set of equations beyond the quasistatic
limit, we find that the model is unstable at large wave numbers $k$
in the past and up to a redshift of order unity. This instability
has been discussed previously by other authors for bimetric models
in general \cite{Comelli:2012db,DeFelice:2014nja} and, if taken at
face value, would rule out the model. Nevertheless, we believe it
is worth analytically identifying the epoch in which the instability
takes place and discussing possible ways to overcome it. This could
help to find other cases, within the class of bimetric models, that
do not suffer from the same problem.

In the regime in which the model is stable we derive its scalar cosmological
perturbation equations in the subhorizon limit and integrate them
numerically. We then compare the results with a recent compilation
of growth data \cite{Macaulay:2013swa}. We find that the MBM fits
both supernovae and growth rate data, while remaining well distinguishable
from $\Lambda$CDM. If a variant of the model is found that cures
the instability in the past, the model could be an interesting competitor
to $\Lambda$CDM.

\section{Background equations}

We start with the action of the form \cite{2012JHEP...02..126H} 
\begin{eqnarray}
S & = & -\dfrac{M_{g}^{2}}{2}\int d^{4}x\sqrt{-g}R(g)-\dfrac{M_{f}^{2}}{2}\int d^{4}x\sqrt{-f}R(f)\\
 & + & m^{2}M_{g}^{2}\int d^{4}x\sqrt{-g}\sum_{n=0}^{4}\beta_{n}e_{n}(X)+\int d^{4}x\sqrt{-g}L_{m}\nonumber 
\end{eqnarray}
where $X_{\gamma}^{\alpha}\equiv\sqrt{g^{\alpha\beta}f_{\beta\gamma}}$,
$e_{n}$ a\textcolor{black}{re elementary symmetric polynomials, $\beta_{n}$}
are arbitrary constants and $L_{m}=L_{m}(g,\psi)$ is a matter Lagrangian.
Here $g_{\mu\nu}$ is the standard metric coupled to matter fields
in the $L_{m}$ Lagrangian, while $f_{\mu\nu}$ is an additional dynamical
tensor field. In the following we express masses in units of the Planck
mass $M_{g}$ and the mass parameter $m^{2}$ will be absorbed into
the parameters $\beta_{n}$. Varying the action with respect to $g{}_{\mu\nu}$,
one obtains the following equations of motion: 
\begin{equation}
G_{\mu\nu}+\dfrac{1}{2}\sum_{n=0}^{3}(-1)^{n}\beta_{n}\left[g_{\mu\lambda}Y_{(n)\nu}^{\lambda}(X)+g_{\nu\lambda}Y_{(n)\mu}^{\lambda}(X)\right]=T_{\mu\nu},\label{eq:eeg}
\end{equation}
where $G_{\mu\nu}$ is Einstein's tensor, and the expressions $Y_{(n)\nu}^{\lambda}(X)$
are defined as 
\begin{align}
Y_{(0)} & =I,\\
Y_{(1)} & =X-I[X],\\
Y_{(2)} & =X^{2}-X[X]+\dfrac{1}{2}I\left([X]^{2}-[X^{2}]\right),\\
Y_{(3)} & =X^{3}-X^{2}[X]+\dfrac{1}{2}X\left([X]^{2}-[X^{2}]\right))\nonumber \\
 & -\dfrac{1}{6}I\left([X]^{3}-3[X][X^{2}]+2[X^{3}]\right),
\end{align}
where $I$ is the identity matrix and $[...]$ is the trace operator.
Varying the action with respect to $f{}_{\mu\nu}$ we get 
\begin{equation}
\bar{G}_{\mu\nu}+\sum_{n=0}^{3}\frac{(-1)^{n}\beta_{4-n}}{2M_{f}^{2}}\left[f_{\mu\lambda}Y_{(n)\nu}^{\lambda}(X^{-1})+f_{\nu\lambda}Y_{(n)\mu}^{\lambda}(X^{-1})\right]=0,
\end{equation}
where the overbar indicates $f{}_{\mu\nu}$ curvatures. Notice that
$\beta_{0}$ acts as a pure cosmological constant. Finally, the rescaling
$f\rightarrow M_{f}^{-2}f$, $\beta_{n}\rightarrow M_{f}^{n}\beta_{n}$
allows us to assume $M_{f}=1$ in the following (see Ref. \cite{2012JCAP...12..021B}).

We assume now a cosmological spatially flat Friedmann-Robertson-Walker
(FRW) metric, 
\begin{equation}
ds^{2}=a^{2}(t)\left(-dt^{2}+dx_{i}dx^{i}\right),
\end{equation}
where $t$ represents the conformal time and a dot will represent
the derivative with respect to it. The second metric is chosen also
in a FRW form 
\begin{equation}
ds_{f}^{2}=-\left[\dot{b}(t)^{2}/\mathcal{H}^{2}(t)\right]dt^{2}+b(t)^{2}dx_{i}dx^{i},
\end{equation}
where $\mathcal{H}\equiv\dot{a}/a$ is the conformal Hubble function.
This form of the metric $f_{\mu\nu}$ ensures that the equations satisfy
the Bianchi identities (see e.g. Ref. \cite{2012JHEP...02..026H}).

Defining $r=b/a$, the background equations can be conveniently written
as a first-order system for $r,\mathcal{H}$, using $N=\log a$ as
the time variable and denoting $d/dN$ with a prime \cite{1475-7516-2014-03-029}
(see also \cite{2013JHEP...03..099A}): 
\begin{align}
2E'E+E^{2} & =a^{2}(B_{0}+B_{2}r'),\\
r' & =\frac{3rB_{1}\Omega_{m}}{\beta_{1}-3\beta_{3}r^{2}-2\beta_{4}r^{3}+3B_{2}r^{2}},\label{eq:rprime}
\end{align}
where $\Omega_{m}=1-\frac{B_{0}}{B_{1}}r$, $E\equiv\mathcal{H}/H_{0}$
and the couplings $\beta_{i}$ are measured in units of $H_{0}^{2}$
and finally 
\begin{align}
B_{0} & =\beta_{0}+3\beta_{1}r+3\beta_{2}r^{2}+\beta_{3}r^{3},\\
B_{1} & =\beta_{1}+3\beta_{2}r+3\beta_{3}r^{2}+\beta_{4}r^{3},\\
B_{2} & =\beta_{1}+2\beta_{2}r+\beta_{3}r^{2}.
\end{align}

\section{Minimal bimetric model}

In Ref. \cite{1475-7516-2014-03-029} we identified the conditions
for standard cosmological viability, i.e. for a matter epoch followed
by a stable acceleration, without bounces or singularities beside
the big bang. We found that among the models with a single nonvanishing
parameter only two cases give a viable cosmology, namely, the cases
with only $\beta_{0}$ or only $\beta_{1}$. The former one is indeed
the $\Lambda$CDM model, while the $\beta_{1}$ case is what we call
the minimal bimetric model. One has then for the MBM 
\begin{equation}
r'=\frac{3r\left(1-3r^{2}\right)}{1+3r^{2}},
\end{equation}
\textcolor{black}{independent of $\beta_{1}$. This equation has two
branches for $r>0$, but only the one that starts at $r=0$ and ends
at $r=1/\sqrt{3}$ is cosmologically viable. In terms of the scale
factor, this solution r}eads \cite{1475-7516-2014-03-029,2013JCAP...12..002F}
\begin{equation}
r(a)=\frac{1}{6}a^{-3}\left(-A\pm\sqrt{12a^{6}+A^{2}}\right),\label{eq:minModel_r}
\end{equation}
where $A=-\beta_{1}+3/\beta_{1}$. \textcolor{black}{These equations
imply a remarkably simple and testable relation between the equation
of state $w$ and $\Omega_{m}$ valid at all times during matter domination:
\begin{equation}
w=\frac{2}{\Omega_{m}-2},\label{eq:testable}
\end{equation}
where the density parameter is given by 
\begin{equation}
\Omega_{m}=1-3r(a)^{2}\;.\label{eq:omega}
\end{equation}
Since the Friedmann equation of the second metric provides $r_{0}=\beta_{1}/3$,
the present value of the matter density parameter is therefore simply
related to single parameter value of the model. Together with Eq.
(\ref{eq:testable}) this shows that all viable parameter values for
$\beta_{1}$ lead to a phantom equation of state at present time.
Another useful relation for the MBM that we will use below is $\mathcal{H}^{2}=\beta_{1}a^{2}/3r$.}

In Ref. \cite{1475-7516-2014-03-029} we found that the MBM fits the
supernovae data well if $\beta_{1}=1.38\pm0.03$, corresponding to
$\Omega_{m0}=1-\beta_{0}^{2}/3=0.37\pm0.02$. The equation of state
turns out to be approximated at small redshifts by $w(z)\approx-1.22_{-0.02}^{+0.02}-0.64_{-0.04}^{+0.05}z/(1+z)$.
However this parametrization is not adequate at $z\ge0.5$ and the
analytic expressions (\ref{eq:minModel_r})-(\ref{eq:omega}) should
be employed instead.

\section{Perturbation equations}

We now find the perturbation equations for the MBM. For the perturbed
part of the metrics we adopt the gauge defined in Fourier space as
\begin{align}
ds_{f}^{2} & =2Fb^{2}\left[-\frac{\dot{b}(t)^{2}\Psi_{f}}{b(t)^{2}\mathcal{H}^{2}(t)}dt^{2}+(\Phi_{f}\delta_{ij}+k_{i}k_{j}E_{f})dx^{i}dx^{j}\right],\nonumber \\
ds^{2} & =2Fa^{2}\left[-\Psi dt^{2}+(\Phi\delta_{ij}+k_{i}k_{j}E)dx^{i}dx^{j}\right],
\end{align}
where $F=e^{i\mathbf{k}\cdot\mathbf{r}}$ . After a transformation
to the gauge-invariant variables \cite{Comelli:2012db} 
\begin{align}
\tilde{\Phi} & =\Phi-\mathcal{H}^{2}E',\\
\tilde{\Psi} & =\Psi-(\mathcal{H}^{2}+\mathcal{HH}')E'-\mathcal{H}^{2}E'',\\
\tilde{\Phi}_{f} & =\Phi_{f}-\frac{r\mathcal{H}^{2}E_{f}'}{(r'+r)},\\
\tilde{\Psi}_{f} & =\Psi_{f}-\frac{\mathcal{H}r^{2}(\mathcal{H}E_{f}')'}{(r'+r)^{2}}-\frac{\mathcal{H}^{2}E_{f}'r(r^{2}+2r'^{2}+2rr'-rr'')}{(r'+r)^{3}},
\end{align}
we obtain from the Einstein equations a set of perturbation equations
in $\Xi=\{\tilde{\Phi},\tilde{\Psi},\tilde{\Phi}_{f},\tilde{\Psi}_{f},E,\Delta E\equiv E-E_{f}\}$,

\begin{align}
[00] & \begin{array}{cc}
\Phi\left(1+\frac{2k^{2}}{3a^{2}r\beta_{1}}\right)-\Phi_{f}+\frac{a^{2}\left(1-3r^{2}\right)\beta_{1}}{-4r+6r^{3}}E'+\frac{\mathcal{H}^{2}\left(1+3r^{2}\right)}{-4+6r^{2}}\text{\ensuremath{\Delta E'}}+\frac{1}{3}k^{2}\text{\ensuremath{\Delta E}}\end{array}\nonumber \\
 & \begin{array}{cc}
-\frac{a\left(-1+3r^{2}\right)\sqrt{\beta_{1}}}{\sqrt{3}k^{2}r^{5/2}}\theta-\frac{\delta\rho}{3B_{2}r}=0,\end{array}\label{eq:pert_g00}\\
{}[0\, i] & \begin{array}{cc}
\Phi'-\Psi+\frac{a^{2}\rho}{2\mathcal{H}k^{2}}\theta+\left(\mathcal{H}^{2}-\mathcal{H}\mathcal{H}'\right)E'=0,\end{array}\label{eq:pert_g0i}\\
{}[i\: j] & \begin{array}{cc}
\Phi+\Psi+a^{2}r\text{\ensuremath{\beta_{1}\Delta E=0}},\end{array}\\
{}[i\,\, i] & \begin{array}{cc}
-\left(2+\frac{2k^{2}}{3a^{2}r\beta_{1}}\right)\Phi+2\Phi_{f}-\Psi\left(1+\frac{2k^{2}}{3a^{2}r\beta_{1}}\right)+\frac{6\left(2-3r^{2}\right)}{3+9r^{2}}\Psi_{f}+\frac{\mathcal{H}^{3}r\left(3+9r^{2}\right)\left(\mathcal{H}-\mathcal{H}'\right)}{a^{2}\left(2-3r^{2}\right)\beta_{1}}E''+\frac{3a^{2}\left(2+9r^{2}\right)\left(1-3r^{2}\right)^{2}\beta_{1}}{4r\left(2-3r^{2}\right)^{2}\left(1+3r^{2}\right)}\ensuremath{E}'\end{array}\nonumber \\
{} & -\frac{2k^{2}}{3}\ensuremath{\Delta E}+\frac{a^{2}\left(1+3r^{2}\right)\beta_{1}}{6r\left(2-3r^{2}\right)}\ensuremath{\Delta E}''+\begin{array}{cc}
\frac{a^{2}\left(22-9r^{2}\left(-19+42r^{2}+15r^{4}\right)\right)\beta_{1}}{12r\left(4-27r^{4}+27r^{6}\right)}\text{\ensuremath{\Delta E'}}=0,\end{array}
\end{align}

\begin{align}
[00] & \begin{array}{cc}
\Phi+\left(-1-\frac{2k^{2}r}{3a^{2}\beta_{1}}\right)\Phi_{f}+\frac{k^{2}}{3}\Delta E+\frac{a^{2}\left(-1+3r^{2}\right)\beta_{1}}{4r-6r^{3}}E'-\frac{a^{2}\left(1+3r^{2}\right)\beta_{1}}{6r\left(2-3r^{2}\right)}\text{\ensuremath{\Delta E'}}=0,\end{array}\\
{}[0\, i] & \begin{array}{cc}
\Phi_{f}'+\frac{\left(-4+6r^{2}\right)}{1+3r^{2}}\Psi_{f}+\frac{3a^{2}\left(-1+3r^{2}\right)\beta_{1}}{4r\left(-2+3r^{2}\right)}E'+\frac{3a^{2}\left(1-3r^{2}\right)\beta_{1}}{4r\left(-2+3r^{2}\right)}\text{\ensuremath{\Delta E'}}=0,\end{array}\label{eq:pert_f0i}\\
{}[i\: j] & \begin{array}{cc}
\Psi_{f}+\Phi_{f}+\frac{a^{2}\left(1+3r^{2}\right)\beta_{1}}{-4r+6r^{3}}\Delta E=0,\end{array}\\
{}[i\,\, i] & \begin{array}{cc}
\Phi+\left(-1+\frac{2k^{2}r\left(-2+3r^{2}\right)}{3a^{2}\left(1+3r^{2}\right)\beta_{1}}\right)\Phi_{f}+\frac{\Psi}{2}+\left(1-\frac{3}{1+3r^{2}}+\frac{2k^{2}r\left(-2+3r^{2}\right)}{3a^{2}\left(1+3r^{2}\right)\beta_{1}}\right)\Psi_{f}+\frac{a^{2}\left(1-3r^{2}\right)\beta_{1}}{4r\left(-2+3r^{2}\right)}E''+\frac{a^{2}\left(1+3r^{2}\right)\beta_{1}}{12r\left(-2+3r^{2}\right)}\text{\ensuremath{\Delta E}}''\end{array}\nonumber \\
{} & \begin{array}{cc}
+\frac{1}{3}k^{2}\text{\ensuremath{\Delta E}}+\frac{a^{2}\left(-22+9r^{2}\left(-19+42r^{2}+15r^{4}\right)\right)\beta_{1}}{24r\left(4-27r^{4}+27r^{6}\right)}\text{\ensuremath{\Delta E}}'-\frac{3\left(2+9r^{2}\right)\left(a-3ar^{2}\right)^{2}\beta_{1}}{8\left(2-3r^{2}\right)^{2}\left(r+3r^{3}\right)}\text{\ensuremath{E}}'=0 & \;,\end{array}\label{eq:pert_fii}
\end{align}
and from the conservation of matter we get two more equations for
the matter density contrast $\delta$ and the velocity divergence
$\theta$,\textcolor{black}{{} 
\begin{equation}
\delta'+\theta\mathcal{H}^{-1}+3\Phi'-3\mathcal{H}^{2}E''-6\mathcal{H}\mathcal{H}'E'+k^{2}E'=0,\label{eq:perteq7_mm}
\end{equation}
\begin{equation}
\theta'+\theta+k^{2}E'\mathcal{H}'-k^{2}\Psi\mathcal{H}^{-1}+k^{2}\mathcal{H}\left(E''+E'\right)=0.\label{eq:perteq8_mm}
\end{equation}
}

\section{Instability}

Recently some authors \cite{Comelli:2012db,DeFelice:2014nja} found
an instability at small scales in massive bimetric theories. Here
we revisit this issue in the MBM\textcolor{black}{. Starting from
the set of general perturbation equations (\ref{eq:pert_g00})-(\ref{eq:pert_fii}),
one can repl}ace all $\Psi_{f},\,\Phi_{f},\,\Delta E$ and their derivatives
by using $g_{00}$, $g_{ii}$ and $g_{ij}$. This also shows that
eqs $g_{ij}$ and $f_{ij}$ are linearly dependent. Then we can replace
$\delta$ and $\theta$ with the help of $g_{0i}$ and $f_{00}$.
Finally, one can find a linear combination of $f_{0i,}$ and $g_{ii}$
which allows one to express $E'$ as a function of $\Psi,\,\Phi$
and their derivatives. In this way, we can express our original ten
equations to just two second-order equations for $X\equiv\{\Psi,\Phi\}$
which can be written as ($i,\, j$=1,2) 
\begin{equation}
X_{i}''+M_{ij}X'_{j}+N_{ij}X_{j}=0,\label{eq:2ndorder-diff-eq}
\end{equation}
where $M_{ij}$ and $N_{ij}$ are two matrices that depend only on
$k$, $\beta_{1}$ and $r$. For the explicit expressions of their
elements see Appendix \ref{sec:Explicit-expression-for-M-and_N}.
The eigenfrequencies of this equation can be found by substituting
$X=X_{0}e^{i\omega N}$, assuming that the dependence of $\omega$
on time is negligibly small. In the limit of large $k$ we find 
\begin{equation}
\omega_{\mp}=\pm\frac{k}{\mathcal{H}}\frac{\sqrt{-1+12r^{2}+9r^{4}}}{1+3r^{2}}
\end{equation}
(here $k$ is in the same units as $\mathcal{H})$ plus two other
solutions, one of which is zero while the is independent of $k$ and
therefore subdominant. One can then see that real solutions (needed
to obtain an oscillating, rather than a growing, solution for $X$)
are found only for $r>0.28$, which occurs for $N\approx-0.4$, i.e.
$z\approx0.5$. This is exactly the same instant at which $r''$ crosses
zero. At any epoch before this, the perturbation equations are unstable
for large $k$, i.e. they grow as $a^{\omega_{+}}$. Notice that $\omega_{\mp}$
are independent of $\beta_{1}$; this means that the instability remains
even in the limit of zero mass, which is similar to the van Dam-Veltman-Zakharov
discontinuity \cite{vanDamVeltman1970,Zakharov1970}. Similar to that
case, one might speculate that when nonlinear order effects start
being important they might cure the instability. Notice also that
the large-$k$ limit we have taken is valid only for $k/\mathcal{H}\gg1$,
i.e. for $r>r_{H}$, where $r_{H}(k)$ is the solution of the equation
$a(r)^{2}=3rk^{2}/\beta_{1}$ and $a(r)$ is obtained by inverting
Eq. (\ref{eq:minModel_r}).

This explosively large growth is in obvious contrast with what we
know about the growth of linear perturbations in our Universe, for
instance, with the smoothness of the microwave cosmic background and
the linearity of present fluctuations on scales larger than a few
megaparsecs. However, one might imagine that by adjusting for instance
the initial conditions or by playing with other assumptions, the model
could be saved. Therefore, in order to quantify the real impact of
the instability, we estimate a directly observable quantity that is
independent of initial conditions: the growth rate of the linear perturbations
as measured with redshift distortions. Since all the perturbation
variables can be written as a linear combination of $\Phi$ and $\Psi$,
their dominant behavior will have the same growth $\sim e^{i\omega_{+}N}$.
This means that during the instability epoch the matter density contrast
grows as $\delta\sim a^{\omega}$ where $\omega=|\omega_{+}|$. This
allows us to estimate the growth rate $f\equiv d\log\delta/dN$ and
to obtain the observable combination $f(z)\sigma_{8}(z)=\sigma_{8}f\delta/\delta_{0}$
as 
\begin{equation}
f(z)\sigma_{8}(z)=Aa^{\omega}(\omega+N\omega'),\label{eq:fs}
\end{equation}
where $A$ is a normalization constant. The combination $f\sigma_{8}(z)$
has been estimated through redshift distortions at various redshifts
up to unity (see for instance Ref. \cite{Macaulay:2013swa}), and
it has been found to be practically constant in the range from $z=0.8$
to $z=0.3$ for scales around $k=0.1h/$Mpc, corresponding to $k/H_{0}\approx50$.
In stark contrast, using the expression $(\ref{eq:fs})$, we estimate
an extremely fast growth during the instability epoch; for instance,
between $z=0.8$ and $z=0.6$ the growth of $f\sigma_{8}(z)$ is found
to be around 180,000 times.

Adding the cosmological constant $\beta_{0}$, one obtains 
\begin{equation}
\omega_{\mp}=\pm\frac{k}{\mathcal{H}}\frac{\sqrt{-1+2(\beta_{0}/\beta_{1})r+12r^{2}+9r^{4}}}{1+3r^{2}}.
\end{equation}
In this case the instability region occurs for any $r<\beta_{1}/2\beta_{0}$;
if $\beta_{1}/\beta_{0}\ll1$ this unstable epoch can be pushed arbitrarily
back into the past but then the model would effectively behave like
$\Lambda$CDM.

It is possible that a different choice of parameters $\beta_{i}$
leads to an evolution which is free from instabilities, or a value
of $r_{H}(k)$ such that (at least for the scales that are today in
the linear regime) the subhorizon evolution occurs during the stable
phase. Finally, one could also assume that $\beta_{1}$ is actually
a time-dependent variable (e.g., it could be a function of a scalar
field, $\beta_{1}(\phi)$), so that its value is very small in the
past - therefore recovering a standard evolution - and comparable
to $H_{0}$ near the present epoch.

\section{Quasistatic limit}

Taken at face value, the instability rules out the MBM, unless nonlinear
effects are able to rescue it. However, we think it is still worthwhile
to consider some of its cosmological effects for two reasons. First,
one of the mentioned mechanisms or some variants thereof might be
able to cure the past instability while leaving unaltered the recent
epoch. Second, the methods we investigate below can be applied to
other choices of parameters in the bimetric class that allow for a
stable evolution.

\begin{figure}
\includegraphics[width=0.7\columnwidth]{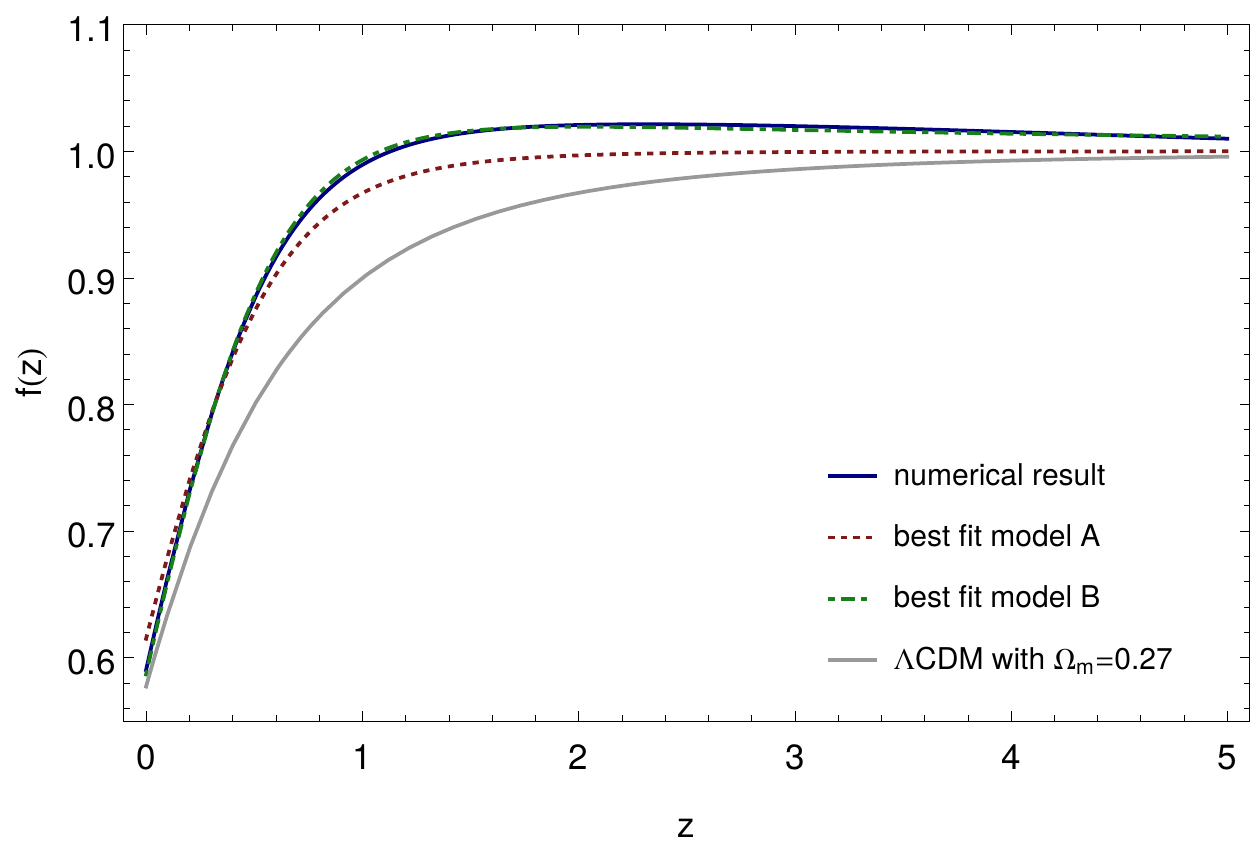}

\protect\protect\caption{Growth rate $f=\delta'/\delta$ in the quasistatic limit for $\beta_{1}=1.39$
and $k=0.1h/\text{Mpc}$. The numerical result (in blue) is approximated
by the fitting model $f=\Omega_{m}^{\gamma}$ (model A, red dotted
curve) and $f=\Omega_{m}^{\gamma_{0}}(1+\frac{\mbox{\ensuremath{\gamma}}_{1}}{z+1})$
(model B, green dash-dotted curve). For a comparison we plot the $\Lambda$CDM
result (gray dashed line) while using $\Omega_{m0}=0.37$ which corresponds
to the present matter density in our analyzed MBM.\label{fig:growthrate}}
\end{figure}

In the regime in which the model is stable, i.e. for $z\le0.5$, one
can simplify the perturbation equations by taking the quasistatic
limit. In this regime and at subhorizon scales, i.e. $k/\mathcal{H}\gg1$,
we can in fact assume that $\Xi_{i}(k/\mathcal{H})^{2}$ is much larger
than $\Xi_{i}$ and its derivatives $\Xi_{i}',\Xi_{i}''$ for any
$\Xi_{i}=\{\Phi,\Psi,\Phi_{f},\Psi_{f},\Delta E,E\}$ and also $\delta(k/\mathcal{H})^{2},\delta'(k/\mathcal{H})^{2}\gg\theta/\mathcal{H}$;
then the set of differential equations becomes algebraic (except for
the matter conservation equations) and we obtain the Poisson-like
relations 
\begin{align}
\Psi & =-\frac{\mathcal{H}^{2}\Omega_{m}\delta\left(2k^{2}r^{3}\left(11+6r^{2}\right)+3\beta_{1}a^{2}\left(1+7r^{2}-6r^{4}\right)\right)}{2k^{2}\left(\beta_{1}a^{2}\left(1+r^{2}\right)^{2}\left(1+3r^{2}\right)+k^{2}r^{3}\left(7+3r^{2}\right)\right)}\;,\\
\Phi & =\frac{\mathcal{H}^{2}\Omega_{m}\delta\left(2k^{2}r^{3}\left(10+3r^{2}\right)+3\beta_{1}a^{2}\left(1+4r^{2}+3r^{4}\right)\right)}{2k^{2}\left(\beta_{1}a^{2}\left(1+r^{2}\right)^{2}\left(1+3r^{2}\right)+k^{2}r^{3}\left(7+3r^{2}\right)\right)}\;,\\
{\color{black}{\color{black}\Psi_{f}}} & {\color{black}{\color{black}={\color{red}{\color{black}-\frac{{\color{black}\mathcal{H}^{2}\Omega_{m}\delta}\left(3r^{2}+1\right)\left(3\beta_{1}a^{2}\left(6r^{4}-7r^{2}-1\right)+2k^{2}r\left(6r^{2}-1\right)\right)}{4k^{2}\left(3r^{2}-2\right)\left(\beta_{1}a^{2}\left(r^{2}+1\right)^{2}\left(3r^{2}+1\right)+k^{2}\left(3r^{2}+7\right)r^{3}\right)}\;,}}}}\\
\Phi_{f} & {\color{black}={\color{red}{\color{black}\frac{\mathcal{H}^{2}\Omega_{m}\delta\left(3r^{2}+1\right)\left(3\beta_{1}a^{2}\left(r^{2}+1\right)+k^{2}r\right)}{2k^{2}\left(\beta_{1}a^{2}\left(r^{2}+1\right)^{2}\left(3r^{2}+1\right)+k^{2}\left(3r^{2}+7\right)r^{3}\right)}\;,}}}\\
\Delta E & ={\color{black}{\color{black}{\color{red}{\color{black}\frac{\mathcal{H}^{2}\Omega_{m}\delta r\left(9\beta_{1}a^{2}\left(1-3r^{2}\right)+2k^{2}r\left(3r^{2}+1\right)\right)}{2a^{2}\beta_{1}k^{2}\left(\beta_{1}a^{2}\left(r^{2}+1\right)^{2}\left(3r^{2}+1\right)+k^{2}\left(3r^{2}+7\right)r^{3}\right)}\;,}}}}
\end{align}
which reduce to the standard ones during the matter epoch, i.e. for
$r\to0$. In the quasistatic limit the set of equations does not contain
the (0,i)$g_{\mu\nu}$ and (0,i)$f_{\mu\nu}$ equations. Since both
equations were used to simplify the remaining ones, we numerically
have checked the consistency of the solutions with both (0,i) equations.
We then obtain the two modified gravity parameters 
\begin{align}
\eta & \equiv-\frac{\Phi}{\Psi}=H_{2}\frac{1+H_{4}(k/\mathcal{H})^{2}}{1+H_{3}(k/\mathcal{H})^{2}},\label{eq:eta}\\
Y & \equiv-\frac{2k^{2}\Psi}{3\mathcal{H}^{2}\Omega_{m}\delta_{m}}=H_{1}\frac{1+H_{3}(k/\mathcal{H})^{2}}{1+H_{5}(k/\mathcal{H})^{2}},\label{eq:Y}
\end{align}
where 
\begin{align}
H_{1} & \equiv\frac{1+7r^{2}-6r^{4}}{\left(1+r^{2}\right)^{2}\left(1+3r^{2}\right)},\\
H_{2} & \equiv\frac{1+4r^{2}+3r^{4}}{1+7r^{2}-6r^{4}},\\
H_{3} & \equiv\frac{2\mathcal{H}^{2}r^{3}\left(11+6r^{2}\right)}{3\text{\ensuremath{\beta_{1}}}a^{2}\left(1+7r^{2}-6r^{4}\right)},\\
H_{4} & \equiv\frac{2\mathcal{H}^{2}r^{3}\left(10+3r^{2}\right)}{3\text{\ensuremath{\beta_{1}}}a^{2}\left(1+4r^{2}+3r^{4}\right)},\\
H_{5} & \equiv\frac{\mathcal{H}^{2}r^{3}\left(7+3r^{2}\right)}{\beta_{1}a^{2}\left(1+r^{2}\right)^{2}\left(1+3r^{2}\right)}.
\end{align}
For $\beta_{1}\to0$ the only consistent background solution is $r\to0$;
in this limit the model reduces to pure CDM and consequently $H_{1,2}=1$
and $H_{3,4,5}=0$. The expressions (\ref{eq:eta}) and (\ref{eq:Y})
have the same structure as the Horndeski Lagrangian \cite{Horndeski:1974,DeFelice:2011hq,2013PhRvD..87b3501A}
since both Lagrangians produce second-order equations of motion. The
matter evolution equations can now be written as a single equation:
\begin{equation}
\delta_{m}''+\delta_{m}'\left(1+\frac{\mathcal{H}'}{\mathcal{H}}\right)-\frac{3}{2}Y(k)\Omega_{m}\delta_{m}=0.\label{eq:delta_subhorizon}
\end{equation}
Integrating numerically this equation along the background solution
(\ref{eq:minModel_r}), we find that near $k=0.1h/\text{Mpc}$ and
$\beta_{1}=1.39$ we can approximate $f\equiv\delta'/\delta\approx\Omega_{m}^{\gamma}$
\cite{Lahav:1991wc} with $\gamma\approx0.47$ in the range $z\in(0,5)$
(see Fig. \ref{fig:growthrate}). Near $\beta_{1}=1.39$ the dependence
on $\beta_{1}$ at $k=0.1h/$Mpc can be linearly approximated as $\gamma=0.26+0.15\beta_{1}$,
while the weak dependence on $k$ is approximately 
\begin{equation}
\gamma(k)=0.47+0.001\left(\frac{k}{0.1h/\mathrm{Mpc}}\right)^{-1/2}\,.\label{eq:kdep}
\end{equation}
Future experiments, like the Euclid satellite \cite{2011arXiv1110.3193L},
plan to measure $\gamma$ to within 0.02; this will amply allow one
to distinguish the MBM from $\Lambda$CDM and standard quintessence,
which predict $\gamma\approx0.54$.

Let us remark, however, that the growth rate is significantly larger
than 1 for redshifts $z\gtrsim1$ and cannot be well approximated
with the standard $\Omega_{m}^{\gamma}$ fit. We find that an additional
correction 
\begin{equation}
f\approx\Omega_{m}^{\gamma_{0}}(1+\frac{\mbox{\ensuremath{\gamma}}_{1}}{z+1})\label{eq:betterfit}
\end{equation}
with $\gamma_{0}=0.58$ and $\gamma_{1}=0.07$ is better able to reproduce
our numerical result.

\begin{figure}
\includegraphics[width=0.7\textwidth]{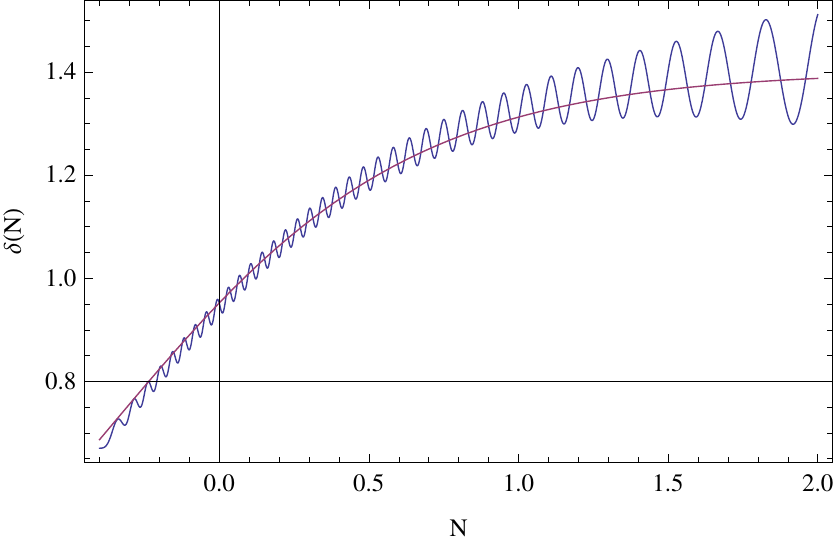}

\protect\protect\caption{The oscillating blue line represents the numerical integration of
the full set of perturbation equations for $k/H_{0}=100,\,\beta_{1}=1.4$.
The red smooth line is the solution of the growth equation (\ref{eq:delta_subhorizon})
in the quasistatic limit for the same parameters.\label{fig:The-oscillating-blue}}
\end{figure}

The quasistatic limit is an excellent approximation to the full behavior,
provided one considers only the stable epoch $z<0.5$, as shown in
Fig. \ref{fig:The-oscillating-blue}.

\section{\textcolor{black}{C}omparison to the growth rate}

The quasistatic limit can be compared to measurements of $f(z)\sigma_{8}(z)$
where $\sigma_{8}(z)=\sigma_{8}G(z)$, with $G(z)$ being the growth
rate normalized to unity today. Most of the present measurements actually
extend to redshifts higher than 0.5, which is outside the stability
regime. Nevertheless, as a way to demonstrate the feasibility of constraining
this model with growth data, we include these measurements as well.
The likelihood is given by 
\begin{equation}
\chi_{f\sigma_{8}}^{2}=\sum_{ij}\left(d_{i}-\sigma_{8}t_{i}\right)C_{ij}^{-1}\left(d_{j}-\sigma_{8}t_{j}\right),
\end{equation}
in which $d_{i}$ and $t_{i}$ are vectors containing the measured
and theoretically expected data, respectively, and $C_{ij}$ denotes
the covariance matrix. Since the current constraints on $\sigma_{8}$
depend on the theory of gravity, for generality we marginalize analytically
the likelihood over $\sigma_{8}>0$, obtaining 
\begin{equation}
\chi_{f\sigma_{8}}^{2}=S_{20}-\frac{S_{11}^{2}}{S_{02}}+\log S_{02}-2\log\left(1+\mathrm{Erf}\left(\frac{S_{11}}{\sqrt{2S_{02}}}\right)\right),\label{chimarg}
\end{equation}
where 
\begin{align}
S_{11} & =d_{i}C_{ij}^{-1}t_{j},\\
S_{20} & =d_{i}C_{ij}^{-1}d_{j},\\
S_{02} & =t_{i}C_{ij}^{-1}t_{j}
\end{align}
(see also e.g. Ref. \cite{Piloyan:2014gta}). Since current data are
not binned in $k$ space, we choose an average value $k=0.1h/$Mpc
in Eq. (\ref{eq:delta_subhorizon}).

\begin{figure}
\textcolor{black}{\includegraphics[width=0.7\columnwidth]{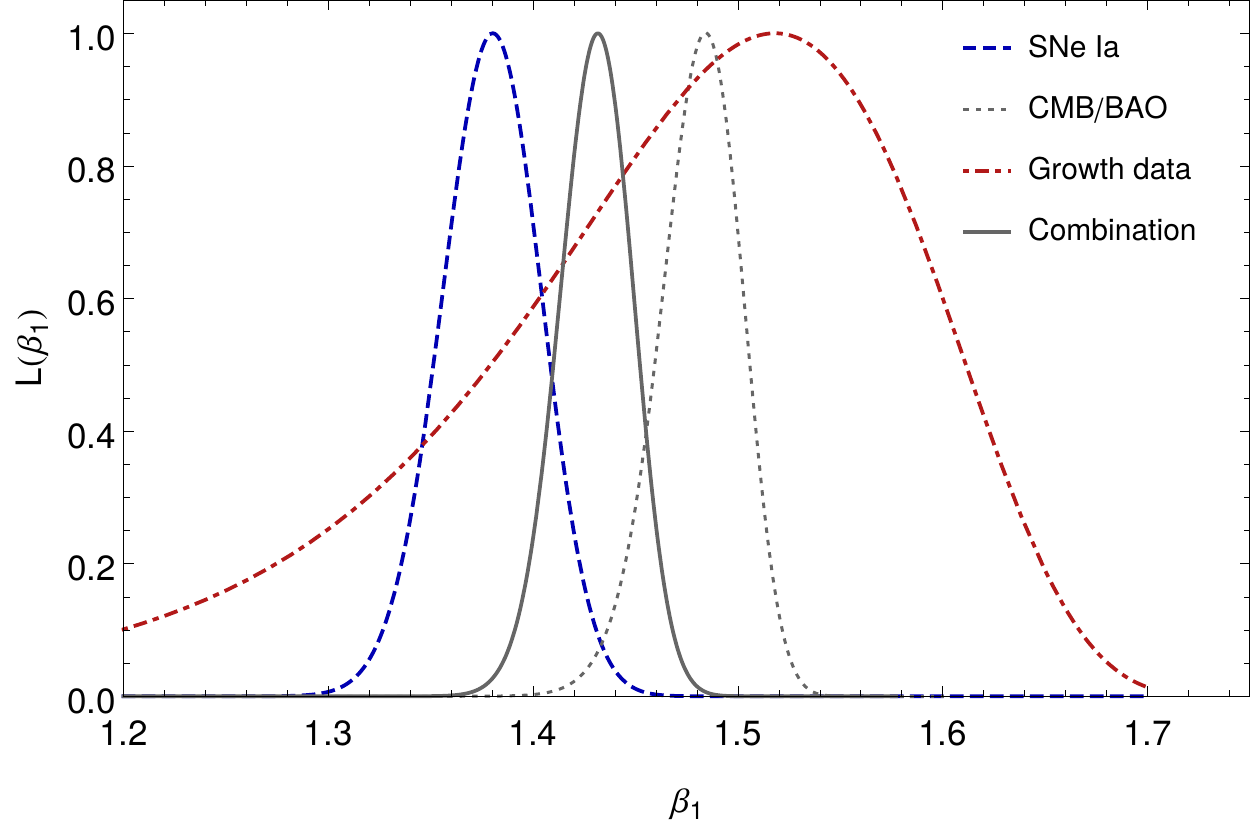}\protect\protect\caption{Likelihood for $\beta_{1}$ obtained from observed SNe Ia (blue dashed),
measured growth data (red dot-dashed) and the combination of CMB and
BAO measurements (dotted gray). The full combined likelihood is indicated
by a gray solid line.\textcolor{red}{{} }All likelihoods are rescaled
to unity at their maximum. For the most likely values we obtain $\beta_{1}=1.38_{-0.03}^{+0.03}$
($\chi_{\text{min}}^{2}=578.3)$ and $\beta_{1}=1.52{}_{-0.13}^{+0.09}$
($\chi_{\text{min}}^{2}=10.48)$ for the comparison with SNe Ia and
growth data, respectively. Due to the broad width of the growth likelihood,
its combination with the other probes does not sensibly change the
results.\label{fig:likelihood}}
}
\end{figure}

We compute the likelihood from the data set compiled in \cite{Macaulay:2013swa}
which contains measured growth histories from the 6dFGS \cite{Beutler:2012px},
LRG$_{200}$, LRG$_{60}$ \cite{Samushia:2011cs}, BOSS \cite{Tojeiro:2012rp},
WiggleZ \cite{Blake:2012pj} and VIPERS \cite{delaTorre:2013rpa}
survey\textcolor{black}{s. }Our results are shown in Fig. \ref{fig:likelihood}.
The growth data constraints appear much broader than, but consistent
with, the supernova type Ia (SN Ia) data. The combined result from
SNe and growth data is $\beta_{1}=1.39\pm0.03$, practically identical
to the best fit from SN Ia alone.\textcolor{red}{{} }We also plot in
Fig. \ref{fig:likelihood} the likelihood from cosmic microwave background
(CMB) and baryon acoustic oscillation (BAO) measurements where we
use the results from the first peak angular size WMAP 7.2 data \cite{Komatsu2011}
and the SDSS DR7 sample including the LRG and 2dFGRS data set \cite{2010MNRAS.401.2148P}.
The combined result from all data, SN + CMB + BAO + growth turns out
to b\textcolor{black}{e $\beta_{1}=1.43\pm0.02$. }However, one should
keep in mind that the CMB data analysis assumes a pure $\Lambda$CDM
so it is not obvious that it can be applied here without corrections.
Note that including the CMB/BAO data does not change the best-fit
parameters for $w(z)$ and $\gamma$ significantly.

\begin{figure}
\textcolor{black}{\emph{\includegraphics[width=0.7\columnwidth]{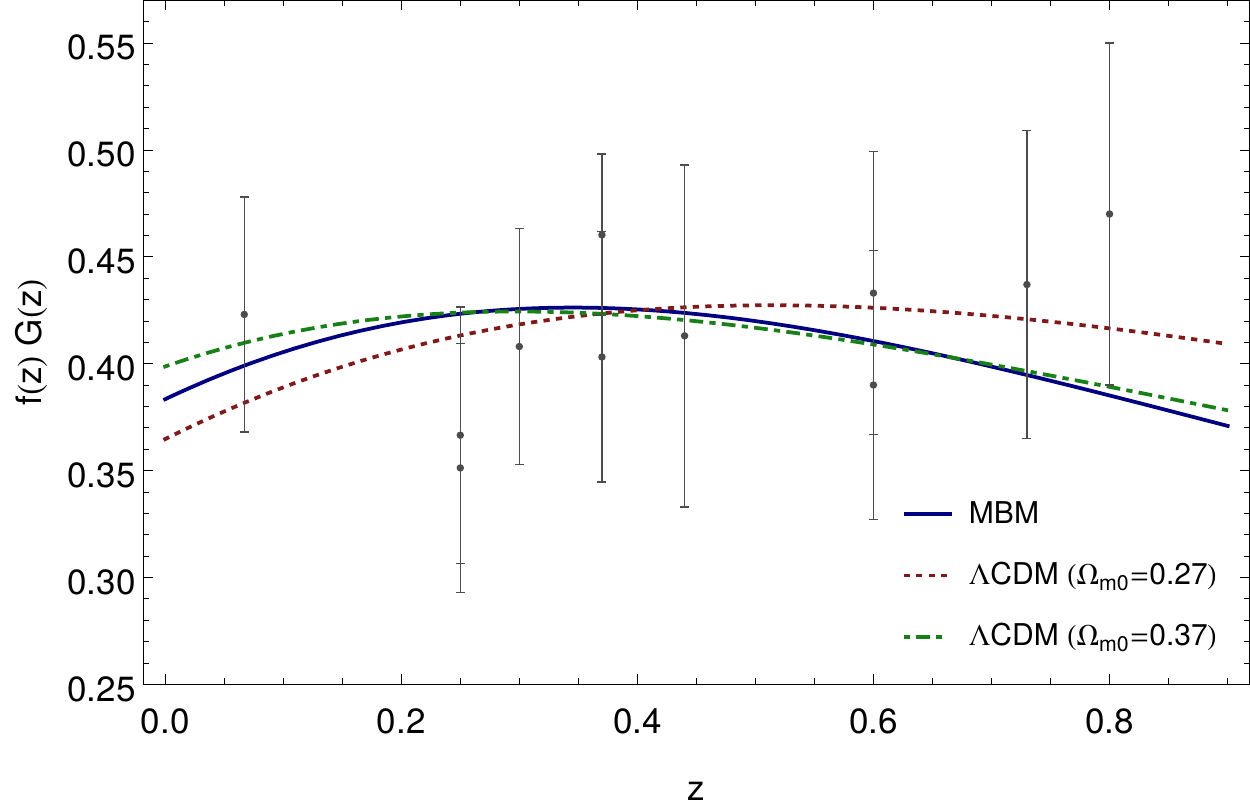}\protect\protect\caption{\textit{\emph{Comparison of growth histories in the MBM with $\beta_{1}=1.39$
(blue) and $\Lambda$CDM (dotted red: $\Omega_{m0}=0.27$; dot-dashed
green: $\Omega_{m0}=0.37$ which corresponds to the present matter
density in the best-fit MBM). The data points are taken from Ref.
\cite{Macaulay:2013swa}. Note that the normalization of the curves
is immaterial due to the marginalization over $\sigma_{8}$.}}\label{fig:growthhistory}}
}}
\end{figure}

Finally, in Fig. \ref{fig:growthhistory} we compare the growth history
corresponding to the most likely MBM with the measured growth data
and the $\Lambda$CDM expectation.

\section{Conclusions}

We have shown that a minimal bimetric model exists which closely reproduces
the success and the simplicity of $\Lambda$CDM at the background
level. We fixed its single parameter, $\beta_{1}$, to percent accuracy
by fitting to supernovae and growth data. The MBM has several unique
signatures, like the $w-\Omega_{m}$ relation (\ref{eq:testable}),
the phantom equation of state, the $k$ dependence of the growth factor
(Eq. \ref{eq:kdep}) and the values of $f$ above unity (Eq. \ref{eq:betterfit}),
all of which will make it easily distinguishable from $\Lambda$CDM
with future experiments.

We have shown however that the model suffers from a perturbation instability
at large $k$ at epochs before $z\approx0.5$, confirming previous
results \cite{Comelli:2012db,DeFelice:2014nja} but also identifying
the exact epoch of transition. Taken at face value, such an instability
seems to rule out this particular form of bimetric model. A possible
way to save the model is to assume that when the perturbations become
nonlinear the instability becomes under control. This conjecture can
be confirmed only by going to higher order in perturbation theory.
Of course the instability might also disappear by choosing a different
set of parameters. We leave a complete analysis of other models to
future work.
\begin{acknowledgments}
We acknowledge support from DFG through the project TRR33 ``The Dark
Universe''. We thank Mariele Motta for discussions and cross-checks
on the perturbation equations and Valerio Marra for help with the
supernovae catalog, and Y. Akrami, T. Koivisto, and A. Solomon for
discussions on the perturbation equations. 
\end{acknowledgments}
\appendix

\section{Explicit expressions for the matrices $M_{ij}$ and $N_{ij}$\label{sec:Explicit-expression-for-M-and_N}}

In this appendix we will present the elements of the matrices $M_{ij}$
and $N_{ij}$ appearing in the second-order differential equation
(\ref{eq:2ndorder-diff-eq}). We start by defining the functions 
\begin{align}
K_{1}\equiv K_{2}\equiv & \frac{4r-6r^{3}}{a^{2}\left(3\beta_{1}r^{2}+\beta_{1}\right)}, & K_{3}\equiv K_{4}\equiv & \frac{1}{a^{2}r\beta_{1}},\\
K_{5}\equiv K_{6}\equiv & -\frac{24k^{2}\left(2-3r^{2}\right)^{2}}{a^{2}r\left(3r^{2}+1\right)^{2}\beta_{1}}, & K_{7}\equiv & -\frac{12\left(2-3r^{2}\right)^{2}\left(3r^{2}-1\right)}{\left(3r^{3}+r\right)^{2}},\\
K_{8}\equiv & -\frac{12\sqrt{3}a\left(2-3r^{2}\right)^{2}\left(3r^{2}-1\right)\sqrt{\beta_{1}}}{r^{5/2}\left(3kr^{2}+k\right)^{2}}, & K_{9}\equiv & -\frac{a\left(3r^{2}-1\right)\sqrt{\beta_{1}}}{k\sqrt{r}},\\
K_{10}\equiv & -\frac{a^{2}k\left(3r^{2}-1\right)\beta_{1}}{\sqrt{3}\left(3r^{3}+r\right)}, & K_{11}\equiv & \frac{4k\left(3r^{2}-2\right)}{\sqrt{3}\left(3r^{2}+1\right)},\\
K_{12}\equiv K_{13}\equiv & \frac{\sqrt{3}a^{2}k\left(3r^{2}-1\right)\beta_{1}}{4r-6r^{3}}, & K_{14}\equiv K_{15}\equiv & \frac{1}{a^{2}r\beta_{1}},\\
K_{16}\equiv & \frac{9r^{2}-6}{k^{2}\left(3r^{2}+1\right)}, & K_{17}\equiv K_{18}\equiv & \frac{1}{a^{2}r\beta_{1}}+\frac{3}{2k^{2}},\\
K_{19}\equiv & \frac{a^{2}\left(135r^{6}+378r^{4}-171r^{2}-22\right)\beta_{1}}{8k^{2}\left(2-3r^{2}\right)^{2}\left(3r^{3}+r\right)}, & K_{20}\equiv & -\frac{9a^{2}\left(1-3r^{2}\right)^{2}\left(9r^{2}+2\right)\beta_{1}}{8k^{2}\left(2-3r^{2}\right)^{2}\left(3r^{3}+r\right)},\\
K_{21}\equiv & \frac{a^{2}\left(3r^{2}+1\right)\beta_{1}}{4k^{2}r\left(3r^{2}-2\right)}, & K_{22}\equiv & -\frac{3a^{2}\left(3r^{2}-1\right)\beta_{1}}{4k^{2}r\left(3r^{2}-2\right)},\\
K_{23}\equiv & \frac{1}{\mathcal{H}}, & K_{24}\equiv & k^{2}-6\mathcal{H}\mathcal{H}',\\
K_{25}\equiv & -3\mathcal{H}^{2}, & K_{26}\equiv & -\frac{k^{2}}{\mathcal{H}},\\
K_{27}\equiv & k^{2}\left(\mathcal{H}+\mathcal{H}'\right), & K_{28}\equiv & k^{2}\mathcal{H},\\
K_{29}\equiv & -\frac{12k^{2}\left(2-3r^{2}\right)^{2}}{\left(3r^{2}+1\right)^{2}}, & K_{30}\equiv & -\frac{36\left(2-3r^{2}\right)^{2}}{\left(3r^{2}+1\right)^{2}},\\
K_{31}\equiv & \frac{12\left(2-3r^{2}\right)^{2}\left(3\beta_{1}a^{2}+2k^{2}r\right)}{a^{2}\left(3r^{2}+1\right)^{2}\beta_{1}}, & K_{32}\equiv & -\frac{6a^{2}\left(3r^{2}-2\right)\beta_{1}}{3r^{3}+r},\\
K_{33}\equiv & \frac{18a^{2}\left(3r^{2}-2\right)\left(3r^{2}-1\right)\beta_{1}}{r\left(3r^{2}+1\right)^{2}}, & K_{34}\equiv & \frac{4r-6r^{3}}{a^{2}\left(3\beta_{1}r^{2}+\beta_{1}\right)},\\
K_{35}\equiv & \frac{4r-6r^{3}}{a^{2}\left(3\beta_{1}r^{2}+\beta_{1}\right)}, & K_{36}\equiv K_{37}\equiv & \frac{1}{a^{2}r\beta_{1}},
\end{align}
which only depend on the background quantities $\beta_{1}$, $r$,
$\mathcal{H}$ and the wave number $k$. Although we introduced several
redundant functions, the definitions of those functions turn out to
be useful since in every bimetric model the dependencies of both $M_{ij}$
and $N_{ij}$ on the $K_{i}$ functions are the same. We proceed bywith
defining{\small{} 
\begin{align}
L_{1}\equiv & K_{13}+\left(3K_{2}K_{31}^{2}\right)^{-1}\left(3K_{1}K_{11}K_{31}K_{33}+2\sqrt{3}kK_{2}\left(K_{33}K_{31}'-K_{31}K_{33}'\right)\right),\\
L_{2}\equiv & \left(3K_{2}K_{31}^{2}\right)^{-1}\left[2\sqrt{3}kK_{2}\left(K_{14}\left(K_{31}\left(K_{29}+K_{32}'\right)-K_{32}K_{31}'\right)-K_{31}\left(K_{30}-2K_{32}K_{14}'\right)\right)\right.\\
 & \left.-3K_{14}K_{31}\left(K_{12}K_{2}K_{31}+K_{1}K_{11}K_{32}\right)\right],\nonumber \\
L_{3}\equiv & \left(3K_{2}K_{31}^{2}\right)^{-1}\left[2\sqrt{3}kK_{2}\left(2K_{31}K_{32}K_{15}'+K_{15}\left(K_{31}\left(K_{29}+K_{32}'\right)-K_{32}K_{31}'\right)\right)\right.\\
 & \left.-3K_{15}K_{31}\left(K_{12}K_{2}K_{31}+K_{1}K_{11}K_{32}\right)\right],\nonumber \\
L_{4}\equiv & -\left(3K_{2}K_{31}^{2}\right)^{-1}\left[3\left(K_{11}K_{3}+K_{12}K_{2}K_{14}'\right)K_{31}^{2}+3K_{1}K_{11}\left(K_{14}K_{29}-K_{30}+K_{32}K_{14}'\right)K_{31}\right.\\
 & \left.-2\sqrt{3}kK_{2}\left(K_{14}K_{29}'-K_{30}'+K_{14}'\left(K_{29}+K_{32}'\right)+K_{32}K_{14}''\right)K_{31}+2\sqrt{3}kK_{2}\left(K_{14}K_{29}-K_{30}+K_{32}K_{14}'\right)K_{31}'\right]\;,\nonumber \\
L_{5}\equiv & \left(3K_{2}K_{31}^{2}\right)^{-1}\left[-3K_{11}K_{4}K_{31}^{2}-3K_{1}K_{11}\left(K_{15}K_{29}+K_{32}K_{15}'\right)K_{31}\right.\\
 & \left.+K_{2}\left(2\sqrt{3}k\left(-K_{32}K_{15}'K_{31}'+K_{29}\left(K_{31}K_{15}'-K_{15}K_{31}'\right)+K_{31}\left(K_{15}K_{29}'+K_{15}'K_{32}'+K_{32}K_{15}''\right)\right)-3K_{12}K_{31}^{2}K_{15}'\right)\right]\;,\nonumber \\
L_{6}\equiv & -\frac{2kK_{33}}{\sqrt{3}K_{31}},\\
L_{7}\equiv & \frac{2kK_{14}K_{32}}{\sqrt{3}K_{31}},\\
L_{8}\equiv & \frac{2kK_{15}K_{32}}{\sqrt{3}K_{31}},\\
L_{9}\equiv & K_{20}+\frac{\left(K_{1}K_{16}k^{2}+3K_{2}\right)K_{33}}{k^{2}K_{2}K_{31}}\;,\\
L_{10}\equiv & K_{14}\left(-K_{19}-\frac{\left(K_{1}K_{16}k^{2}+3K_{2}\right)K_{32}}{k^{2}K_{2}K_{31}}\right)-2K_{21}K_{14}',\\
L_{11}\equiv & K_{15}\left(-K_{19}-\frac{\left(K_{1}K_{16}k^{2}+3K_{2}\right)K_{32}}{k^{2}K_{2}K_{31}}\right)-2K_{21}K_{15}',\\
L_{12}\equiv & -\left(k^{2}K_{2}K_{31}\right)^{-1}\left[-K_{18}K_{2}K_{31}k^{2}+K_{16}K_{3}K_{31}k^{2}+K_{1}K_{16}\left(K_{14}K_{29}-K_{30}+K_{32}K_{14}'\right)k^{2}\right.\\
 & \left.+K_{19}K_{2}K_{31}K_{14}'k^{2}+K_{2}K_{21}K_{31}K_{14}''k^{2}-3K_{2}K_{30}+K_{14}K_{2}\left(K_{31}k^{2}+3K_{29}\right)+3K_{2}K_{32}K_{14}'\right],\nonumber \\
L_{13}\equiv & K_{17}-K_{19}K_{15}'-K_{21}K_{15}''-\left(k^{2}K_{2}K_{31}\right)^{-1}\left[K_{16}K_{31}K_{4}k^{2}\right.\\
 & \left.+K_{1}K_{16}\left(K_{15}K_{29}+K_{32}K_{15}'\right)k^{2}+K_{15}K_{2}\left(K_{31}k^{2}+3K_{29}\right)+3K_{2}K_{32}K_{15}'\right],\nonumber \\
L_{14}\equiv & K_{22},\\
L_{15}\equiv & -K_{14}K_{21},\\
L_{16}\equiv & -K_{15}K_{21},\\
L_{17}\equiv & \left[K_{9}\left(K_{7}\left(K_{24}K_{7}K_{9}K_{31}^{2}+\left(K_{31}K_{8}K_{10}'+K_{6}K_{9}K_{33}'+K_{33}K_{9}K_{6}'\right)K_{31}-K_{33}K_{6}K_{9}K_{31}'\right)-K_{31}K_{33}K_{6}K_{9}K_{7}'\right)\right.\\
 & \left.-K_{10}K_{31}^{2}\left(K_{9}\left(K_{23}K_{7}^{2}-K_{8}'K_{7}+K_{8}K_{7}'\right)+K_{7}K_{8}K_{9}'\right)\right]\left(K_{31}K_{7}K_{9}\right)^{-2},\nonumber \\
L_{18}\equiv & \left(\sqrt{3}K_{31}K_{7}K_{9}\right)^{-2}\left[-2\sqrt{3}kK_{31}^{2}\left(K_{9}\left(K_{23}K_{7}^{2}-K_{8}'K_{7}+K_{8}K_{7}'\right)+K_{7}K_{8}K_{9}'\right)+3K_{9}^{2}\left(K_{14}K_{31}K_{32}K_{6}K_{7}'\right.\right.\\
 & \left.\left.-K_{7}\left(\left(K_{5}-3K_{7}\right)K_{31}^{2}+\left(K_{6}\left(-K_{30}+2K_{32}K_{14}'+K_{14}\left(K_{29}+K_{32}'\right)\right)+K_{14}K_{32}K_{6}'\right)K_{31}-K_{14}K_{32}K_{6}K_{31}'\right)\right)\right]\;,\nonumber \\
L_{19}\equiv & \left(3K_{31}^{2}K_{7}^{2}K_{9}\right)^{-1}\left[-2\sqrt{3}kK_{7}K_{8}K_{31}^{2}+3K_{15}K_{32}K_{6}K_{7}K_{9}K_{31}'\right.\\
 & \left.+3K_{9}\left(K_{15}K_{32}K_{6}K_{7}'-K_{7}\left(2K_{32}K_{6}K_{15}'+K_{15}\left(K_{6}\left(K_{29}+K_{32}'\right)+K_{32}K_{6}'\right)\right)\right)K_{31}\right],\nonumber \\
L_{20}\equiv & -\left(K_{31}K_{7}\right)^{-2}\left[K_{7}K_{5}'K_{31}^{2}-K_{5}K_{7}'K_{31}^{2}+K_{14}K_{6}K_{7}K_{29}'K_{31}-K_{6}K_{7}K_{30}'K_{31}+K_{6}K_{7}K_{14}'K_{32}'K_{31}\right.\\
 & \left.-K_{30}K_{7}K_{6}'K_{31}+K_{32}K_{7}K_{14}'K_{6}'K_{31}+K_{30}K_{6}K_{7}'K_{31}-K_{32}K_{6}K_{14}'K_{7}'K_{31}+K_{32}K_{6}K_{7}K_{14}''K_{31}\right.\nonumber \\
 & \left.+K_{30}K_{6}K_{7}K_{31}'-K_{32}K_{6}K_{7}K_{14}'K_{31}'+K_{29}\left(K_{31}\left(K_{6}K_{7}K_{14}'+K_{14}K_{7}K_{6}'-K_{14}K_{6}K_{7}'\right)-K_{14}K_{6}K_{7}K_{31}'\right)\right],\nonumber \\
L_{21}\equiv & \left(\sqrt{3}K_{31}K_{7}K_{9}\right)^{-2}\left[2\sqrt{3}kK_{31}^{2}\left(K_{9}\left(K_{23}K_{7}^{2}-K_{8}'K_{7}+K_{8}K_{7}'\right)+K_{7}K_{8}K_{9}'\right)\right.\\
 & \left.-3K_{9}^{2}\left(K_{15}K_{31}K_{6}K_{7}K_{29}'+K_{29}\left(K_{31}\left(K_{6}K_{7}K_{15}'+K_{15}K_{7}K_{6}'-K_{15}K_{6}K_{7}'\right)-K_{15}K_{6}K_{7}K_{31}'\right)\right.\right.\nonumber \\
 & \left.\left.+K_{15}'\left(K_{31}K_{6}K_{7}K_{32}'+K_{32}\left(K_{31}K_{7}K_{6}'-K_{6}\left(K_{7}K_{31}'+K_{31}K_{7}'\right)\right)\right)+K_{31}K_{32}K_{6}K_{7}K_{15}''\right)\right],\nonumber \\
L_{22}\equiv & K_{25}+K_{7}^{-1}\left(\frac{K_{33}K_{6}}{K_{31}}+\frac{K_{10}K_{8}}{K_{9}}\right),\\
L_{23}\equiv & \left(3K_{7}\right)^{-1}\left(\frac{2\sqrt{3}kK_{8}}{K_{9}}-\frac{3K_{14}K_{32}K_{6}}{K_{31}}\right),\\
L_{24}\equiv & -\frac{K_{15}K_{32}K_{6}}{K_{31}K_{7}},\\
L_{25}\equiv & K_{26}+\frac{2k\left(K_{9}-K_{9}'\right)}{\sqrt{3}K_{9}^{2}},\\
L_{26}\equiv & \frac{K_{10}K_{9}'-K_{9}\left(K_{10}-K_{27}K_{9}+K_{10}'\right)}{K_{9}^{2}},\\
L_{27}\equiv & \frac{2k\left(K_{9}'-K_{9}\right)}{\sqrt{3}K_{9}^{2}},\\
L_{28}\equiv & \frac{2k}{\sqrt{3}K_{9}},\\
L_{29}\equiv & K_{28}-\frac{K_{10}}{K_{9}},\\
L_{30}\equiv & -\frac{2k}{\sqrt{3}K_{9}},
\end{align}
}and 
\begin{align}
G_{1}\equiv & \frac{-L_{12}L_{23}L_{8}+L_{12}L_{24}L_{7}+L_{15}L_{20}L_{8}-L_{15}L_{24}L_{4}-L_{16}L_{20}L_{7}+L_{16}L_{23}L_{4}}{-L_{1}L_{15}L_{24}+L_{1}L_{16}L_{23}+L_{15}L_{17}L_{8}-L_{16}L_{17}L_{7}-L_{23}L_{8}L_{9}+L_{24}L_{7}L_{9}},\\
G_{2}\equiv & \frac{-L_{13}L_{23}L_{8}+L_{13}L_{24}L_{7}+L_{15}L_{21}L_{8}-L_{15}L_{24}L_{5}-L_{16}L_{21}L_{7}+L_{16}L_{23}L_{5}}{-L_{1}L_{15}L_{24}+L_{1}L_{16}L_{23}+L_{15}L_{17}L_{8}-L_{16}L_{17}L_{7}-L_{23}L_{8}L_{9}+L_{24}L_{7}L_{9}},\\
G_{3}\equiv & \frac{-L_{10}L_{23}L_{8}+L_{10}L_{24}L_{7}+L_{15}L_{18}L_{8}-L_{15}L_{2}L_{24}-L_{16}L_{18}L_{7}+L_{16}L_{2}L_{23}}{-L_{1}L_{15}L_{24}+L_{1}L_{16}L_{23}+L_{15}L_{17}L_{8}-L_{16}L_{17}L_{7}-L_{23}L_{8}L_{9}+L_{24}L_{7}L_{9}},\\
G_{4}\equiv & \frac{-L_{11}L_{23}L_{8}+L_{11}L_{24}L_{7}+L_{15}L_{19}L_{8}-L_{15}L_{24}L_{3}-L_{16}L_{19}L_{7}+L_{16}L_{23}L_{3}}{-L_{1}L_{15}L_{24}+L_{1}L_{16}L_{23}+L_{15}L_{17}L_{8}-L_{16}L_{17}L_{7}-L_{23}L_{8}L_{9}+L_{24}L_{7}L_{9}}.
\end{align}

The elements of the matrices $M_{ij}$ and $N_{ij}$ are then given
by 
\begin{align}
M_{11}= & \frac{L_{11}-G_{4}L_{9}-L_{14}\left(G_{2}+G_{4}'\right)+\left(L_{23}-G_{3}L_{22}\right)^{-1}\left(G_{3}L_{14}-L_{15}\right)\left(-G_{4}L_{17}+L_{19}-L_{22}\left(G_{2}+G_{4}'\right)\right)}{-G_{4}L_{14}+L_{16}+\left(L_{23}-G_{3}L_{22}\right)^{-1}\left(G_{3}L_{14}-L_{15}\right)\left(L_{24}-G_{4}L_{22}\right)},\\
M_{12}= & \frac{L_{10}-G_{3}L_{9}-L_{14}\left(G_{1}+G_{3}'\right)+\left(L_{23}-G_{3}L_{22}\right)^{-1}\left(G_{3}L_{14}-L_{15}\right)\left(-G_{3}L_{17}+L_{18}-L_{22}\left(G_{1}+G_{3}'\right)\right)}{-G_{4}L_{14}+L_{16}+\left(L_{23}-G_{3}L_{22}\right)^{-1}\left(G_{3}L_{14}-L_{15}\right)\left(L_{24}-G_{4}L_{22}\right)},\\
M_{21}= & \left(L_{16}\left(L_{23}-G_{3}L_{22}\right)+L_{15}\left(G_{4}L_{22}-L_{24}\right)+L_{14}\left(G_{3}L_{24}-G_{4}L_{23}\right)\right)^{-1}\left[L_{14}L_{17}G_{4}^{2}-L_{22}L_{9}G_{4}^{2}-L_{14}L_{19}G_{4}\right.\\
 & \left.+L_{24}L_{9}G_{4}+L_{11}\left(G_{4}L_{22}-L_{24}\right)+G_{2}L_{14}L_{24}+L_{14}L_{24}G_{4}'+L_{16}\left(-G_{4}L_{17}+L_{19}-L_{22}\left(G_{2}+G_{4}'\right)\right)\right],\nonumber \\
M_{22}= & \left(L_{16}\left(G_{3}L_{22}-L_{23}\right)+L_{15}\left(L_{24}-G_{4}L_{22}\right)+L_{14}\left(G_{4}L_{23}-G_{3}L_{24}\right)\right)^{-1}\left[-G_{3}G_{4}L_{14}L_{17}+G_{4}L_{14}L_{18}\right.\\
 & \left.-G_{1}L_{14}L_{24}+L_{10}\left(L_{24}-G_{4}L_{22}\right)+G_{3}G_{4}L_{22}L_{9}-G_{3}L_{24}L_{9}-L_{14}L_{24}G_{3}'+L_{16}\left(G_{3}L_{17}-L_{18}+L_{22}\left(G_{1}+G_{3}'\right)\right)\right]\nonumber 
\end{align}
and 
\begin{align}
N_{11}= & \frac{L_{13}-G_{2}L_{9}-L_{14}G_{2}'+\left(L_{23}-G_{3}L_{22}\right)^{-1}\left(G_{3}L_{14}-L_{15}\right)\left(-G_{2}L_{17}+L_{21}-L_{22}G_{2}'\right)}{-G_{4}L_{14}+L_{16}+\left(L_{23}-G_{3}L_{22}\right)^{-1}\left(G_{3}L_{14}-L_{15}\right)\left(L_{24}-G_{4}L_{22}\right)},\\
N_{12}= & \frac{L_{12}-G_{1}L_{9}-L_{14}G_{1}'+\left(L_{23}-G_{3}L_{22}\right)^{-1}\left(G_{3}L_{14}-L_{15}\right)\left(-G_{1}L_{17}+L_{20}-L_{22}G_{1}'\right)}{-G_{4}L_{14}+L_{16}+\left(L_{23}-G_{3}L_{22}\right)^{-1}\left(G_{3}L_{14}-L_{15}\right)\left(L_{24}-G_{4}L_{22}\right)},\\
N_{21}= & \left(L_{16}\left(L_{23}-G_{3}L_{22}\right)+L_{15}\left(G_{4}L_{22}-L_{24}\right)+L_{14}\left(G_{3}L_{24}-G_{4}L_{23}\right)\right)^{-1}\left[G_{2}G_{4}L_{14}L_{17}-G_{4}L_{14}L_{21}\right.\\
 & \left.+L_{13}\left(G_{4}L_{22}-L_{24}\right)-G_{2}G_{4}L_{22}L_{9}+G_{2}L_{24}L_{9}+L_{14}L_{24}G_{2}'+L_{16}\left(-G_{2}L_{17}+L_{21}-L_{22}G_{2}'\right)\right],\nonumber \\
N_{22}= & \left(L_{16}\left(L_{23}-G_{3}L_{22}\right)+L_{15}\left(G_{4}L_{22}-L_{24}\right)+L_{14}\left(G_{3}L_{24}-G_{4}L_{23}\right)\right)^{-1}\left[G_{1}G_{4}L_{14}L_{17}-G_{4}L_{14}L_{20}\right.\\
 & \left.+L_{12}\left(G_{4}L_{22}-L_{24}\right)-G_{1}G_{4}L_{22}L_{9}+G_{1}L_{24}L_{9}+L_{14}L_{24}G_{1}'+L_{16}\left(-G_{1}L_{17}+L_{20}-L_{22}G_{1}'\right)\right].\nonumber 
\end{align}

\vfill{}

\medskip{}

 \bibliographystyle{luc2}
\bibliography{observables,massive-gravity,amendola}

\end{document}